\documentclass[twocolumn,pre]{revtex4}
\usepackage{natbib}
\usepackage{definitions}
\usepackage{localdefs}
\usepackage{amsmath}
\usepackage{amssymb}
\usepackage{graphicx}
\usepackage{amsbsy}
\usepackage{bm}

\begin{document}
\date{\today}

\title{Equilibrium and nonequilibrium thermodynamics of particle-stabilized
thin liquid films}

\author{J.\ B{\l}awzdziewicz}
\affiliation{Department of Mechanical Engineering, Yale University,
P.O. Box 20-8286, New Haven, CT 06520}
\author{E.\ Wajnryb}
\affiliation{IPPT, \'Swi\c{e}tokrzyska 21, Warsaw, Poland}  

\begin{abstract}

Our recent quasi-two-dimensional thermodynamic description of
thin-liquid films stabilized by colloidal particles is generalized to
describe nonuniform equilibrium states of films in external potentials
and nonequilibrium transport processes produced in the film by
gradients of thermodynamic forces.  Using a Monte--Carlo simulation
method, we have determined equilibrium equations of state for a film
stabilized by a suspension of hard spheres.  Employing a
multipolar-expansion method combined with a flow-reflection technique,
we have also evaluated the short-time film-viscosity coefficients and
collective particle mobility.

\end{abstract}

\maketitle

\section{introduction}

\enlargethispage{4pt}

Owing to the presence of oscillatory structural forces, static and
dynamic properties of thin liquid films stabilized by colloidal
particles, micelles, or macromolecules are quite different than the
properties of particle-free films 
\cite{%
Kralchevsky-Danov-Ivanov:1996,
Bergeron:1999,
Stubenrauch-von_Klitzing:2003%
}.  
A striking example of thin-film
behavior that is caused by oscillatory structural forces is the
stepwise-thinning (i.e., film stratification) phenomenon.  Stepwise
thinning---or thickening
\cite{Czarnecki-Taylor-Masliyah:2005}---occurs in liquid films
stabilized by colloidal particles
\cite{%
Nikolov-Wasan-Kralchevsky-Ivanov:1992,%
Nikolov-Wasan:1992,%
Basheva-Danov-Kralchevsky:1997,%
Henderson-Trokhymchuk-Nikolov-Wasan:2005%
},
polyelectrolytes
\cite{
Bergeron-Radke:1992,%
Theodoly-Tan-Ober-Williams-Bergeron:2001%
},
and surfactant solutions \cite{Tchoukov-Mileva-Exerova:2004}.  Such films
undergo a sequence of stepwise transitions between regions of different, but
uniform, thickness. In some cases, quite complex multiphase structures are
obtained \cite{Horiuchi-Matsumara-Furusawa:1998,Czarnecki:2005}.

The stepwise-thinning in particle-stabilized thin liquid films has
been studied experimentally
\cite{%
Nikolov-Wasan:1992,%
Kralchevsky-Danov-Ivanov:1996,%
Basheva-Danov-Kralchevsky:1997,%
Sethumadhavan-Nikolov-Wasan:2001a,%
Sethumadhavan-Bindal-Nikolov-Wasan:2002,%
Basheva-Kralchevsky-Danov-Ananthapadmanabhan-Lips:2007%
}
and theoretically
\cite{%
Kralchevsky-Danov-Ivanov:1996,%
Chu-Nikolov-Wasan:1995,%
Trokhymchuk-Henderson-Nikolov-Wasan:2001,%
Trokhymchuk-Henderson-Nikolov-Wasan:2003,%
Henderson-Trokhymchuk-Nikolov-Wasan:2005,%
Blawzdziewicz-Wajnryb:2005%
}.
Most of the analyses of the mechanism of stepwise thinning focused on
the role of {\it normal} structural force produced by the suspended
particles
\cite{%
Kralchevsky-Danov-Ivanov:1996,%
Chu-Nikolov-Wasan:1995,%
Trokhymchuk-Henderson-Nikolov-Wasan:2001,%
Trokhymchuk-Henderson-Nikolov-Wasan:2003,%
Henderson-Trokhymchuk-Nikolov-Wasan:2005,%
Basheva-Kralchevsky-Danov-Ananthapadmanabhan-Lips:2007%
}.  
However, as we have pointed out in our recent paper
\cite{Blawzdziewicz-Wajnryb:2005}, not only the {\it normal} but also {\it
lateral} structural forces play an important role in the film-stratification
phenomenon. 

To include the lateral structural forces in our description of phase
equilibria in particle-stabilized thin liquid films we have proposed a
quasi-two-dimensional thermodynamic formalism
\cite{Blawzdziewicz-Wajnryb:2005,Previous_descriptions_note}.
\nocite{Kralchevsky-Danov-Ivanov:1996,%
Ikeda-Krustev-Muller:2004,%
Stubenrauch-von_Klitzing:2003}
In this formalism the film area, film volume, and number of colloidal
particles are independent parameters of state.  The corresponding
conjugate intensive parameters are the film tension, normal pressure,
and chemical potential of the colloidal particles in the film.

The film tension involves the {\it anisotropic} part of the
osmotic-pressure tensor of the suspension confined between the film
interfaces.  Hence, the lateral component of the structural force
produced by the suspended particles contributes to mechanical
equilibrium in the lateral direction (i.e., direction along the film).
An analysis of the lateral structural forces is thus essential for
describing film properties, in particular, for formulating the
equilibrium coexistence conditions between film phases of different
thickness \cite{Blawzdziewicz-Wajnryb:2005}.

In the present paper the ideas introduced in
\cite{Blawzdziewicz-Wajnryb:2005} are developed in two directions.
First, we propose a theoretical description of equilibrium states of
films with varying thickness $h$ and nonuniform particle number
density per unit area $\nCarea$ when lateral external forces are
present.  Second, we present a systematic analysis of macroscopic film
dynamics, based on effective 2D transport equations that involve
viscous dissipation and diffusive particle flux.

We also provide numerical results for equilibrium and nonequilibrium
properties of a film stabilized by a suspension of hard spheres.  The
normal stress and film tension are obtained from the contact value of
the pair distribution, which is calculated using a Monte--Carlo
technique.  To determine the chemical potential of the particles
confined in the film we propose a new evaluation method, based on
integration of the Gibbs-Duhem relation that links variations of the
particle chemical potential to the corresponding variations of the
normal osmotic pressure and film tension.  The transport coefficients
in our nonequilibrium theory are determined using a
Stokesian-dynamics algorithm for a periodic system of spheres
\cite{Cichocki-Felderhof-Hinsen-Wajnryb-Blawzdziewicz:1994,%
Ekiel_Jezewska-Wajnryb:2008}, 
combined with a flow-reflection technique.

Our paper is organized as follows.  A general framework for a
description of nonuniform equilibrium states in particle-stabilized
thin-liquid films is developed in Sec.\ \ref{Equilibrium description},
followed by an analysis of nonequilibrium transport processes in
Sec.\ \ref{Macroscopic transport properties}.  Short-time transport
coefficients are evaluated in Sec.\ \ref{Evaluation of short-time
transport coefficients}, and concluding remarks are given in Sec.\
\ref{Conclusions}.

\section{Equilibrium description}
\label{Equilibrium description}

\subsection{Particle-stabilized films}
\label{The system}

We consider a thin liquid film of thickness $h$ and area $A$
stabilized by a colloidal suspension of hard spheres of diameter
$d\sim h$.  The film interfaces are surfactant-free and have a
constant interfacial tension $\sigma$.  The film is surrounded by an
inert gas of a constant pressure $\externalP$.  The viscosity and mass
density of the gas are much smaller than the corresponding parameters
for the fluid in the film; therefore, the gas interacts with the film
only through the static pressure $\externalP$.  The temperature $T$ in
the system is uniform.

We assume that the film is approximately planar.  The film thickness
and the number of particles per unit area are either constant or
slowly varying in the lateral directions $x$ and $y$ on the length
scale $L\gg h$.  Under these conditions the transverse suspension
structure relaxes on the time scale $\tStruct$ that is much shorter
than the time scale $\tLat$ for evolution of the long-wavelength
lateral modes.  After the time $\tStruct$, the suspension in the film
is thus in a local-equilibrium state, characterized by densities per
unit area of conserved quantities such as the volume or number of
colloidal particles.

Three-dimensional macroscopic quantities that appear in our analysis
(e.g., stress and pressure tensors, suspension velocity, and particle
flux) are the ensemble averages of the corresponding microscopic
quantities.  Assuming that the film properties are slowly varying in
the lateral directions, the local area averages and ensemble averages
are equivalent.  In the effective quasi-two-dimensional description of
the film, the averages across the film of 3D quantities are evaluated,
in addition to the ensemble averages.

In all our numerical examples the suspending fluid is treated as an
incompressible continuous medium.  The fluid affects the thermodynamic
state of the system only through the isotropic contribution to the
pressure tensor and through the constant-volume constraint.  The
particles in the film interact via the hard-sphere potential.  We
note, however, that our thermodynamic considerations apply more
generally, e.g., to suspensions of charged particles or small
particles with finite-size hydration layers.  In our hydrodynamical
calculations the creeping-flow conditions are assumed, and many-body
hydrodynamic interactions of the particles in the film are fully taken
into account.

\begin{figure}
\includegraphics*{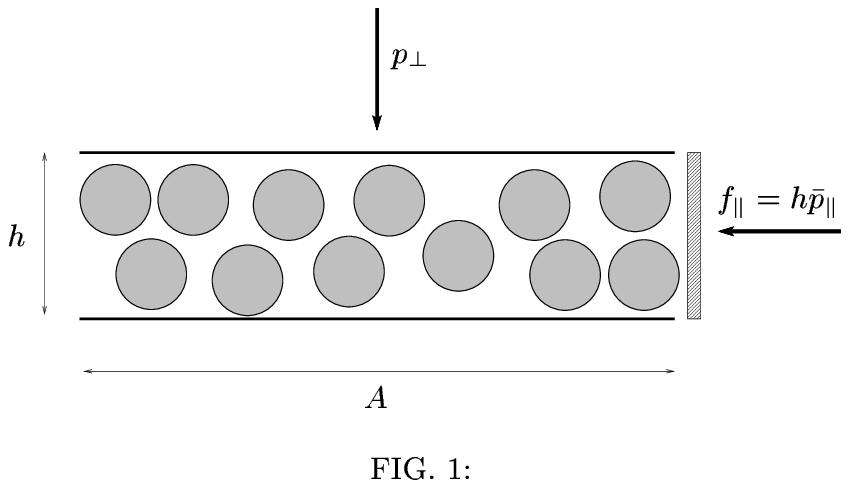}
\caption{Geometry of a particle-stabilized thin liquid film.}
\label{Definition figure}
\end{figure}

\subsection{Normal pressure and film tension}
\label{Normal pressure and film tension}

Unlike the corresponding quantity in bulk suspensions, the equilibrium
pressure tenor $\Ptensor$ in particle-stabilized films is
anisotropic.  This anisotropy stems from different particle
ordering in the transverse and lateral directions.  For films with
parallel (or nearly parallel) planar interfaces, the pressure tensor
has only the normal and lateral components,
\begin{equation}
\label{pressure tensor}
\Ptensor=\normalP\ez\ez+\lateralP\lateralUnitTensor,
\end{equation}
owing to the axial symmetry of the problem.  Here
$\lateralUnitTensor=\ex\ex+\ey\ey$ is the lateral unit tensor (where
$\ex$ and $\ey$ are unit vectors in the lateral directions $x$ and
$y$), and $\ez$ is the unit vector in the transverse direction $z$
(normal to the film).  

Due to continuity of the momentum flux, the normal pressure $\normalP$
is independent of the transverse coordinate $z$.  However, the lateral
component $\lateralP$ of the pressure tensor depends on the transverse
position in the film.  In our paper the average value across the film
of a quantity $b$ (such as the lateral stress $\lateralP$) will be
denoted by the overbar, i.e.,
\begin{equation}
\label{average across film}
\bar b=h^{-1}\int_0^h\diff z\,b(z).
\end{equation}
Normal and lateral stresses acting on a section of the film are schematically
illustrated in Fig.\ \ref{Definition figure}.

The infinitesimal work associated with the change of the film
thickness $h$ and area $A$ is 
\begin{equation}
\label{work}
\dd W=-A\normalP\diff h+(2\sigma-h\lateralPaver)\diff A.
\end{equation}
This expression includes the pressure and surface-tension
contributions.

The quantity $f_\parallel=(2\sigma-h\lateralPaver$) in Eq.\
\refeq{work} represents the lateral force (per unit length of a film
section) resulting from the interfacial-tension and pressure
contributions.  This force, however, cannot be interpreted as film
tension, and a correct relation for the film tension is given in Eq.\
\refeq{film tension in terms of stress components}.  The term
$(2\sigma-h\lateralPaver)\diff A$ in Eq.\ \refeq{work} describes the
work associated with a change of the film area at a constant {\it
thickness} $h$.  The description in terms of independent variables $h$
and $A$ would give meaningless results for a film composed of an
incompressible suspension, because any area change in such a system
must be accompanied by the corresponding thickness change.  Moreover,
since we have $\lateralPaver=\normalP$ for a film composed of a simple
isotropic fluid, the force $f_\parallel$ depends on the external
pressure $\externalP$ (which is in equilibrium with the normal
pressure $\normalP$ in the film).  The film tension, however, should
be pressure independent.

A correct film-tension representation of the equilibrium states of the
film is obtained by rewriting the pressure tensor \refeq{pressure
tensor} in the form
\begin{equation}
\label{pressure tensor rewritten}
\Ptensor=\normalP\unitTensor+(\lateralP-\normalP)\lateralUnitTensor,
\end{equation}
where $\unitTensor=\lateralUnitTensor+\ez\ez$ is the 3D unit tensor.
The first term on the right-hand side of the above relation is an
isotropic pressure component that should not contribute to the film
tension (as discussed above).  The second term, integrated across the
gap, corresponds to the {\it excess\/} lateral force per unit length
of film section.  Only this excess force contributes to the film
tension $\tension$.  Combining the excess lateral pressure with the
surface-tension contribution we obtain the relation for the film
tension,
\begin{equation}
\label{film tension in terms of stress components}
\tension=h(\normalP-\lateralPaver)+2\sigma
\end{equation}
and the corresponding expression for the work,
\begin{equation}
\label{work in terms of film tension}
\dd W=-\normalP\diff V+\tension\diff A,
\end{equation}
where $V=Ah$ is the film volume.  The term $\tension\diff A$ in the
above equation describes the work associated with the change of the
film area at a constant film {\it volume}, in contrast to the
corresponding term in Eq.\ \refeq{work}.

In mechanical equilibrium, the normal pressure in the film equals the
constant pressure of the surrounding gas,
\begin{equation}
\label{pressure equilibrium condition}
\normalP=\externalP.
\end{equation}
The excess lateral stresses have to balance as well, which implies
that the film tension is position independent
\begin{equation}
\label{film tension equilibrium condition}
\tension=\const,
\end{equation}
provided that there are no external forces acting on the film. 

In the above discussion we assumed that there is no direct interaction
between the film interfaces.  However, if there is such an interaction
force (e.g., the van der Waals attraction or electrostatic repulsion),
the normal pressure $\normalP$ in Eqs.\ \refeq{work}--\refeq{pressure
equilibrium condition} should be replaced with
$\normalP'=\normalP+f_{\rm n}$, where $f_{\rm n}$ is the normal force
per unit area, and $f_{\rm n}<0$ corresponds to attraction.

\subsection{Quasi-two-dimensional thermodynamic description}
\label{Quasi-two dimensional thermodynamic description}

Owing to the time scale separation $\tStruct\ll\tLat$, the relaxation
of the suspension structure in the transverse direction is much faster
than the relaxation of long-wavelength lateral modes.  In a
quasi-two-dimensional thermodynamic description
\cite{Blawzdziewicz-Wajnryb:2005} all details of the transverse
structure of the film are thus averaged out.  The film thickness
$h=V/A$, however, is retained as a thermodynamic variable, because it
can be controlled by varying the external pressure.

The fundamental thermodynamic relation for the free energy $F$ in the
film-tension representation can be obtained from the expression for
work \refeq{work in terms of film tension} by using the standard
entropy-maximization arguments \cite{Callen:1985}.  Treating the
suspension in the film as a two-component fluid (with the colloidal
particles regarded as macromolecules) we get the fundamental relation
in the free-energy representation
\begin{equation}
\label{fundamental relation for film free energy}
\diff F=-S\diff T-\normalP\diff V+\tension\diff A
       +\muC\diff \NC+\muF\diff \NF,
\end{equation}
where $S$ is the entropy, and $\NC$ and $\NF$ denote, respectively, the
number of colloidal particles and solvent molecules in the system, and
$\muC$ and $\muF$ are the corresponding chemical potentials.
According to the above relation and the free-energy-minimum principle,
a particle-stabilized film in thermodynamic equilibrium must satisfy
the thermal equilibrium condition $T=\const$, the mechanical
equilibrium conditions \refeq{pressure equilibrium condition} and
\refeq{film tension equilibrium condition}, and the chemical
equilibrium condition \refstepcounter{equation}
\label{chemical equilibrium conditions}
\newcommand{\chemEQ}{\theequation}
\begin{equation}
\tag{\arabic{equation}a,b}
\muC=\const, \qquad\muF=\const.
\end{equation}

A direct consequence of the fundamental relation \refeq{fundamental
relation for film free energy} is the Gibbs phase rule for coexisting
regions of different thickness $h$ in stratified films in
thermodynamic equilibrium.  Assuming that the suspending fluid behaves
as incompressible continuum medium, its degrees of freedom can be
neglected (as discussed in Sec.\ \ref{Excess quantities}).  Thus, the
state of each phase is described by three densities per unit area: the
excess entropy density, the film thickness $h=V/A$ and the number
density of the colloidal particles $\nCarea=\NC/A$.  There are also
four equilibrium conditions corresponding to the intensive parameters
$T$, $\normalP$, $\tension$, and $\muC$.  It follows that a two-phase
system has $f=2$ degrees of freedom, and the maximum number of
coexisting phases is $k=4$.

We note that in experiments with vertical films, up to seven
coexisting phases of different thickness were observed
\cite{Basheva-Danov-Kralchevsky:1997}.  In such systems, additional
thermodynamic degrees of freedom are provided by the external gravity
potential, and it is possible that not all equilibrium conditions are
satisfied (due to dynamical constraints resulting in slow relaxation
of certain intensive-parameter differences
\cite{Blawzdziewicz-Wajnryb:2005}).  

\subsection{Nonuniform systems}
\label{Nonuniform systems}

\subsubsection{External forces}

The physical meaning of film tension \refeq{film tension in terms of
stress components} can be further elucidated by considering nonuniform
equilibrium states of a film in external lateral force fields.  In
Sec.\ \ref{Macroscopic transport properties} an analysis of such
nonuniform film states will also help to determine the linear
constitutive relations governing film dynamics.

We consider potential forces\refstepcounter{equation}
\label{two force fields a,b}
\newcommand{\EQforceFields}{\theequation}
\begin{equation}
\tag{\arabic{equation}a,b}
\bfFluid=-\bnablaS\psiFluid,\qquad\bfColl=-\bnablaS\psiColl,
\end{equation}
acting, respectively, on the fluid molecules and colloidal particles.
It is assumed that the potentials $\psiFluid$ and $\psiColl$ depend
only on the lateral position $\brho=x\ex+y\ey$.  

The force fields \refeq{two force fields a,b} are normalized per
single particle.  Thus, the total force density per unit area of the
film is
\begin{equation}
\label{force per unit surface}
\bFs=
  \nFarea\bfFluid+\nCarea\bfColl,
\end{equation}
where $\nFarea$ and $\nCarea$ are the local values of the area number
densities $\nFarea=\NF/A$ and $\nCarea=\NC/A$ of the fluid molecules
and colloidal particles at the lateral position $\brho$.  Similarly,
the total force density per unit volume is given by the relation
\begin{equation}
\label{total volume force}
\bfVolTot=\nFvol\bfFluid+\nCvol\bfColl,
\end{equation}
where $\nFvol$ and $\nCvol$ are the local number densities per unit volume of
the fluid molecules and colloidal particles at a (three-dimensional) position
$\br$.  

\subsubsection{Lateral stress balance}

The balance between the film tension and external force \refeq{force per unit
surface} can be obtained from the full 3D stress-balance
equation
\begin{equation}
\label{pressure balance equation}
\bnabla\bcdot\Ptensor=\bfVolTot
\end{equation}
(which is valid in the whole space).  In the region inside the film, the
pressure tensor $\Ptensor$ is given by Eq.\ \refeq{pressure tensor rewritten};
outside the film it is equal to the external pressure tensor,
\begin{equation}
\label{pressure outside film}
\Ptensor=\externalP\unitTensor.
\end{equation}
In Eq.\ \refeq{pressure balance equation}, the interfacial
contribution is neglected under the assumptiona that there is no
spacial variation of the interfacial tension, $\sigma=\const$,
and that the film thickness varies on the
length scale $L\gg h$ (and thus the film curvature is negligible).

Combining relations \refeq{pressure tensor rewritten} and
\refeq{pressure outside film} for the pressure tensor inside and
outside the film, and using the normal-stress boundary condition
\refeq{pressure equilibrium condition}, we find that the divergence of
the isotropic pressure component vanishes in the whole space.  The
integral of the remaining part with respect to the variable $z$ yields
the lateral stress balance (cf.\ Appendix \ref{Effective lateral
stress})
\begin{equation}
\label{integral for tension balance}
\bnablaS\bcdot h(\lateralPaver-\normalP)\lateralUnitTensor=\bFs,
\end{equation}
where $\bFs$ is the area force density \refeq{force per unit surface}.
By using definition \refeq{film tension in terms of stress components}
of film tension, and applying the assumptions $L\gg h$ and
$\sigma=\const$ to include the interfacial-tension term, the lateral
stress-balance equation \refeq{integral for tension balance} can be
expressed as the hydrostatic condition for the film-tension gradient
\begin{equation}
\label{tension balance}
\bnablaS\tension=-\bFs.
\end{equation}

Equation \refeq{tension balance} requires that, in equilibrium, the
area force density be a gradient of a 2D potential,
 \begin{equation}
\label{potential of area force}
\bFs=-\bnablaS\areaPotential.
\end{equation}
Thus, in the presence of external force field \refeq{force per unit
surface}, the equilibrium condition \refeq{film tension equilibrium
condition} is replaced by the hydrostatic condition for the film
tension,
\begin{equation}
\label{nonuniform lateral mechanical equilibrium}
\tension-\areaPotential=\const.
\end{equation}
A more general discussion of the lateral stress balance in the film,
applicable to equilibrium and nonequilibrium states, is given in
Appendix \ref{Effective lateral stress}.

\subsubsection{Chemical potentials}

External forces \refeq{two force fields a,b} modify not only the
lateral stress balance, but also the equilibrium conditions
\refeq{chemical equilibrium conditions} for the chemical potentials of
the suspending fluid and colloidal component.  As in the standard
thermodynamics of bulk systems, the modified chemical equilibrium
conditions are \refstepcounter{equation}
\label{chemical equilibrium conditions nonuniform}
\begin{equation}
\tag{\arabic{equation}a,b}
\muF+\psiFluid=\const, \qquad\muC+\psiColl=\const.
\end{equation}
We note that the above expressions are consistent with the lateral
stress balance \refeq{tension balance}, owing to the Gibbs--Duhem
relation
\begin{equation}
\label{Gibbs--Duhem general}
-S\diff T+V\diff\normalP-A\diff\tension
       -\NC\diff\muC-\NF\diff\muF=0
\end{equation}
(which is obtained from \refeq{fundamental relation for film free energy} by
the usual Legendre transformation of all extensive variables).
At constant $T$ and $\normalP$, the Gibbs--Duhem  relation yields
\begin{equation}
\label{nonuniform system - consistency with Gibbs--Duhem}
\bnablaS\tension=
       -\nCarea\bnablaS\muC-\nFarea\bnablaS\muF.
\end{equation}
By inserting the equilibrium conditions \refeq{chemical equilibrium
conditions nonuniform} into \refeq{nonuniform system - consistency
with Gibbs--Duhem} and using the definition \refeq{force per unit
surface} of the area force density $\bFs$, the lateral stress-balance
equation \refeq{tension balance} is recovered.

\subsection{Incompressible suspension}
\label{Excess quantities}

In our further considerations we focus on films composed of an
incompressible suspension.  It is assumed that both the solvent and
the suspended particles are incompressible, and changes of the
suspension volume due to interaction of particle hydration layers are
negligible.

The number of solvent molecules in such a film is set by the film
volume $V$ and the colloidal-particle volume fraction
$\volumeFractionC=\frac{1}{6}\pi d^3 \nCvolAver$, where
$\nCvolAver=\NC/V$ is the particle number density averaged over film
thickness.  The chemical potential $\muF$ of the solvent is not an
independent function of state, either.  Under isothermal conditions,
assumed herein, $\muF$ is fully determined by the Gibbs--Duhem
relation for the suspending fluid
\begin{equation}
\label{Gibbs--Duhem for fluid}
\nFFvol\diff\muF=\diff\pFluid
\end{equation}
(where $\nFFvol$ is the number density per unit volume for the fluid
molecules in pure solvent).  The thermodynamic degrees of freedom
associated with the solvent variables $\muF$ and $\NF$ can thus be
eliminated from the fundamental relation \refeq{fundamental relation
for film free energy}.

The thermodynamic description of an incompressible system can be
rephrased in terms of appropriate excess quantities describing the
colloidal contributions to thermodynamic functions of state.  In this
section we define excess quantities and formulate the corresponding
equilibrium conditions.

The pressure tensor in the film,
\begin{equation}
\label{pressure decomposition into the fluid and colloid contributions}
\Ptensor=\pFluid\unitTensor+\PCtensor,
\end{equation}
 is decomposed into the isotropic fluid pressure $\pFluid$ and the
osmotic pressure tensor
\begin{equation}
\label{colloid pressure tensor}
\PCtensor=\normalPC\ez\ez+\lateralPC\lateralUnitTensor
\end{equation}
that results from the presence of the colloidal particles.  The
isotropic fluid pressure $\pFluid$ does not contribute to film
tension, according to the definition \refeq{film tension in terms of
stress components}.  It is, therefore, convenient to rewrite the film
tension
\begin{equation}
\label{tension as surface tension and excess terms}
\tension=\tensionC+2\sigma
\end{equation}
as a sum of the interfacial tension term and the colloidal
contribution
\begin{equation}
\label{colloidal contribution to tension}
\tensionC=h(\normalPC-\lateralPCaver).
\end{equation}
Since we assume no variation of the interfacial tension along the
film, the colloidal contribution $\tensionC$ differs from the full
film tension $\tension$ only by a constant.  Thus, the
hydrostatic-equilibrium condition for film tension \refeq{nonuniform
lateral mechanical equilibrium} remains, essentially, unchanged,
\begin{equation}
\label{equilibrium condition for excess film tension}
\tensionC-\areaPotential=\const.
\end{equation}

In a system with no external potentials, the fluid pressure satisfies
the equilibrium condition $\pFluid=\const$.  In the presence of an
external potential \refeqa{two force fields a,b}{a}, the
chemical-equilibrium condition \refeqa{chemical equilibrium conditions
nonuniform}{a} and the Gibbs-Duhem relation \refeq{Gibbs--Duhem for
fluid} yield
\begin{equation}
\label{hydrostatic condition}
\pFluid+\psiFluidVol=\const,
\end{equation}
where
\begin{equation}
\label{volume potential acting on fluid}
\psiFluidVol=\nFFvol\psiFluid
\end{equation}
is the potential of the volume force acting on the fluid phase.
Equation \refeq{hydrostatic condition} represents the usual
hydrostatic-equilibrium condition for the suspending fluid.

With the help of the hydrostatic relation \refeq{hydrostatic
condition}, the constant-normal-pressure condition \refeq{pressure
equilibrium condition} can be transformed into the corresponding
equilibrium condition for the osmotic normal pressure $\normalPC$,
\begin{equation}
\label{equilibrium conditions for excess pressure}
   \normalPC-\psiFluidVol=\const.
\end{equation}
The equilibrium stress-balance equation \refeq{equilibrium conditions
for excess pressure} relates the normal osmotic pressure in the film
to the potential $\psiFluidVol$ of the lateral force acting on the
fluid.  Thus, in equilibrium states with a nonzero lateral force, the
normal osmotic pressure varies along the film.

The remaining equilibrium condition that has to be considered, is
relation \refeqa{chemical equilibrium conditions nonuniform}{b} for
the chemical potential of the colloidal particles $\muC$.  In the
fundamental relation \refeq{fundamental relation for film free
energy}, the term $\muC\diff\NC$ corresponds to the change of the free
energy $F$ when a colloidal particle is added to the system at a fixed
volume and number of fluid particles.  However, in an incompressible
suspension, a particle that is moved from one subsystem of the film to
another is always replaced by the equivalent amount of the suspending
fluid.  Accordingly, the quantity $\muC$ involves not only the
configurational contribution resulting from the distribution of the
colloidal particles, but also an additional part corresponding to the
amount of the fluid displaced by a particle.

To define the excess chemical potential $\muCexcess$ that is free of
the fluid contribution, we begin with the Gibbs--Duhem relation
\refeq{Gibbs--Duhem general}.  By combining \refeq{Gibbs--Duhem
general} with the Gibbs--Duhem relation for the suspending fluid
\refeq{Gibbs--Duhem for fluid} we obtain the identity
\begin{equation}
\label{modified Gibbs--Duhem}
V\diff\normalP-A\diff\tension
       -\NC\diff\muC-(1-\volumeFractionC)V\diff\pFluid=0,
\end{equation}
where the constant-temperature conditions are assumed.  Defining the excess
chemical potential of the colloidal particles as
\begin{equation}
\label{excess chemical potential}
\muCexcess=\muC-\vParticle\pFluid
\end{equation}
(where $\vParticle$ is the volume of a colloidal particle), and using
relation $\normalP=\normalPC+\pFluid$, we find the Gibbs--Duhem
relation for $\muCexcess$,
\begin{equation}
\label{excess Gibbs--Duhem relation}
\diff\muCexcess=\nCvolAver^{-1}\diff\normalPC-\nCarea^{-1}\diff\tensionC.
\end{equation}
The above equation indicates that the quantity $\muCexcess$ is
determined (up to a temperature-dependent additive constant) by the
osmotic normal pressure and film tension, which justifies our
interpretation of $\muCexcess$ as the excess chemical potential.

In order to rephrase the chemical equilibrium condition for the
colloidal particles in terms of the excess chemical potential
$\muCexcess$, relations \refeq{hydrostatic condition} and
\refeqa{chemical equilibrium conditions nonuniform}{b} are inserted
into \refeq{excess chemical potential}.  As the result we get the
chemical equilibrium condition for the excess quantities
\begin{equation}
\label{equilibrium condition for excess chemical potential}
\muCexcess+\psiCollExcess=\const,
\end{equation}
where the excess potential 
\begin{equation}
\label{colloidal potential with buoyancy}
\psiCollExcess=\psiColl-\vParticle\psiFluidVol
\end{equation}
of the external force acting on the colloidal particles includes the
buoyancy contribution resulting from the pressure gradient in the
suspending fluid.

\subsection{Numerical results}
\label{Numerical results equilibrium}

\subsubsection{Evaluation technique}
\label{Evaluation technique}

In this section we present some numerical results for the
osmotic-pressure tensor $\PCtensor$, film tension $\tensionC$, and
particle chemical potential $\muCexcess$ for a film stabilized by a
suspension of hard spheres of diameter $d$.  A standard Metropolis
Monte--Carlo algorithm was used to generate equilibrium ensemble for a
system of a hundred hard spheres in a film with periodic boundary
conditions in the lateral directions.  The results were obtained for
several volume fractions of colloidal particles $\volumeFractionC$, as
a function of film thickness $h$.

\paragraph{Osmotic-pressure tensor}

The ensemble-averaged osmotic pressure tensor has been evaluated using
a hard-sphere generalization of the Kirkwood--Buff expression for the
pressure tensor in an inhomogeneous fluid \cite{Kirkwood-Buff:1949}.
Accordingly, the colloidal contribution to the pressure tensor
$\PCtensor$, averaged over the space $\hat V=A(h-d)$ accessible to the
particle centers, is obtained from the relation
\begin{equation}
\label{particle contribution to stress}
\frac{\averPrimPCtensor}{kT}=\nCvolAverPrime
   +\halff d^3\,\left\langle\int \hat \br\hat \br \,
     n_2^\eq(\br_1,\br_1-d\hat \br)\,\diff^2\hat r\right\rangle_{\hat V}.
\end{equation}
Here $n_2^\eq(\br_1,\br_2)$ is the two-particle equilibrium reduced
distribution, $\hat\br=\br/r$ is the unit vector along the line
passing through the particle centers, the integration is over the
contact configurations, $\langle\cdots\rangle_{\hat V}$ denotes the
volume average over the region $\hat V$, and $\nCvolAverPrime=\NC/\hat
V$ is the corresponding particle number density.

Equation \refeq{particle contribution to stress} generalizes a
well-known expression for the pressure in a bulk hard-sphere system in
terms of the contact value of the radial distribution.  The
contact-value relation \refeq{particle contribution to stress} can be
derived either from the collisional contribution to the momentum flux
in a hard-sphere system, or by passing to the hard-sphere limit in the
Kirkwood-Buff expression \cite{Kirkwood-Buff:1949} for the pressure
tensor in an inhomogeneous fluid.

In our numerical calculations, the average \refeq{particle
contribution to stress} is evaluated from a sum of contributions
originating from pairs of particles separated by a distance
$r_{ij}<d(1+\epsilon)$, where $\epsilon\ll1$.  The results for finite
values of $\epsilon$ are extrapolated to obtain the limiting result
for $\epsilon\to0$.

The contact-value relation \refeq{particle contribution to stress}
provides a convenient basis for numerical evaluation of the pressure
tensor in the film.  However, $\averPrimPCtensor$ involves only
averaging over the accessible volume $\hat V$, and does not
incorporate the momentum transfer from the particle centers to the
wall.  To obtain the pressure tensor $\averPCtensor$ averaged over the
whole volume of the film $V$, we need to consider this momentum
transfer.

The normal momentum transfer occurs when a particle collides with the
interface \cite{collision_note}.  Thus, there is a constant normal
momentum flux in the geometrically excluded regions of thickness
$a=\halff d$ next to each interface.  It follows that the normal
component of the pressure is not affected by the choice of the region
($V$ or $\hat V$) over which the average is performed.  In contrast
there is no lateral momentum transport in the excluded layers next to
each wall, and the averages over the volumes $V$ and $\hat V$ differ
by the factor $\hat V/V=(h-d)/h$.

The above arguments imply that the normal and lateral components of
the tensor $\PCtensor$ averaged over the volume $V$ are
\begin{equation}
\label{corrected pressure}
\normalPC=\normalPCaverPrime,\qquad
\lateralPCaver=h^{-1}(h-d)\lateralPCaverPrime,
\end{equation}
where $\averPrimPCtensor = \normalPCaverPrime\ez\ez +
\lateralPCaverPrime\lateralUnitTensor$ is given by the contact-value
relation \refeq{particle contribution to stress}.  The components
\refeq{corrected pressure} are used to determine the colloidal
contribution to film tension \refeq{colloidal contribution to
tension}.

\paragraph{Chemical potential}

Our quasi-two-dimensional thermodynamic formalism can be conveniently
used for evaluation of the chemical potential of the colloidal
particles in the film $\muCexcess$ from the results for the normal
pressure and film tension. Our method relies on numerical integration
of the Gibbs--Duhem relation \refeq{excess Gibbs--Duhem relation},
using results of Monte--Carlo simulations at different values of
particle volume fraction $\volumeFractionC$ and film thickness $h$.

For convenience, in our calculations we use a modified Gibbs--Duhem
relation based directly on the pressure tensor \refeq{particle
contribution to stress}, without introducing the averages
\refeq{corrected pressure} over the full film volume $V$.  In this
modified formulation, the Gibbs--Duhem relation \refeq{excess
Gibbs--Duhem relation} is replaced with
\begin{equation}
\label{modified excess Gibbs--Duhem relation}
\diff\muCexcess
   =\nCvolAverPrimeInv\diff\normalPC-\nCarea^{-1}\diff\tensionCprime,
\end{equation}
where the modified film tension $\tensionCprime$ is given by the
formula
\begin{equation}
\label{modified tension}
\tensionCprime=(h-d)(\normalPC-\lateralPCaverPrime),
\end{equation}
instead of \refeq{colloidal contribution to tension}. 

We note that equations \refeq{excess Gibbs--Duhem relation} and
\refeq{modified excess Gibbs--Duhem relation} are equivalent, which
follows from the identities
\begin{equation}
\label{relation between different tension representations}
\tensionC=\tensionCprime+d\normalPC,
\end{equation}
\begin{equation}
\label{relation between different densities}
\nCvolAver^{-1}=\nCvolAverPrimeInv+d\nCarea^{-1},
\end{equation}
where, as previously, the overbar and the hat indicate the average
particle density in the volumes $V$ and $\hat V$, respectively, and
the tilde indicates the particle density per unit area.

We have verified that numerical integration of Gibbs-Duhem relation
\refeq{modified excess Gibbs--Duhem relation} over different paths in
the thermodynamic space of states yields equivalent results.  To
obtain proper normalization of the chemical potential, relation
\refeq{modified excess Gibbs--Duhem relation} is integrated from a
low-density state, where we can use the perfect-gas result for the
chemical potential.

Since equilibrium properties of a confined suspension do not depend on
the details of the dynamic boundary conditions at the interfaces of
the film (e.g., free interface or rigid wall), our method for
evaluating the chemical potential can be applied not only to thin
liquid films, but also to other systems with a slit-pore geometry.  In
particular, our method can be used to determine the chemical potential
of spherical macromolecules in the gap between two much larger
particles in a system where depletion interactions are important.  Our
method can also be generalized to systems interacting via continuous
or anisotropic potentials.

\begin{figure}

\includegraphics*{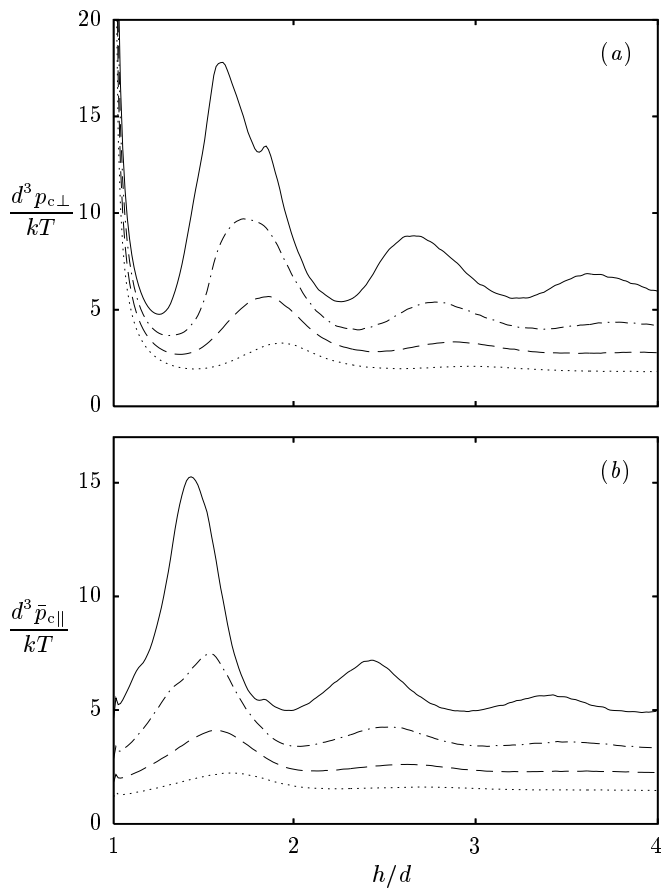}

\caption{
\subfig{a} Normal and \subfig{b} lateral osmotic pressure in the film,
versus film thickness $h$, for particle volume fractions
$\volumeFractionC=0.25$ (dotted line), 0.3 (dashed), 0.35
(dot--dashed), and 0.4 (solid).
}
\label{pressures versus h}
\end{figure}

\subsubsection{Results}

\begin{figure}

\includegraphics*{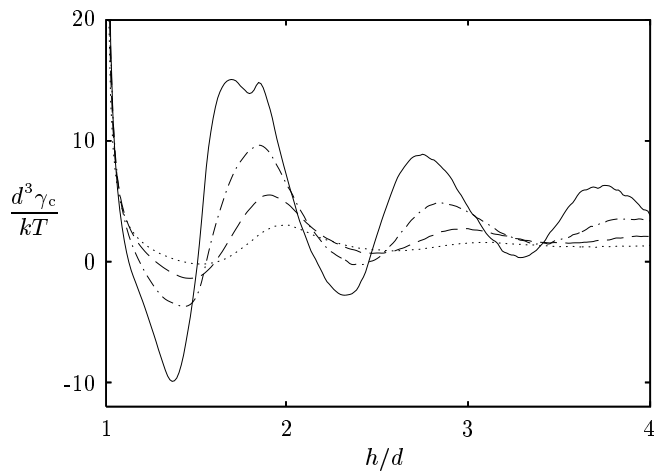}

\caption{
Particle contribution to film tension, versus film thickness $h$, for
the same values of volume fraction $\volumeFractionC$ as in Fig.~\ref{pressures versus h}.
}
\label{film tension versus h}
\end{figure}

The dependence of the normal and lateral pressure components
\refeq{corrected pressure} on the film thickness is illustrated in
Fig.\ \ref{pressures versus h}, which shows $\normalPC$ and
$\lateralPCaver$ versus $h$ for several values of the particle volume
fraction $\volumeFractionC$.  The corresponding behavior of the film
tension $\tensionC$ and excess chemical potential $\muCexcess$ is
depicted in Figs.\ \ref{film tension versus h} and \ref{chemical
potential versus h} \cite{tension_note}.

\begin{figure}

\includegraphics*{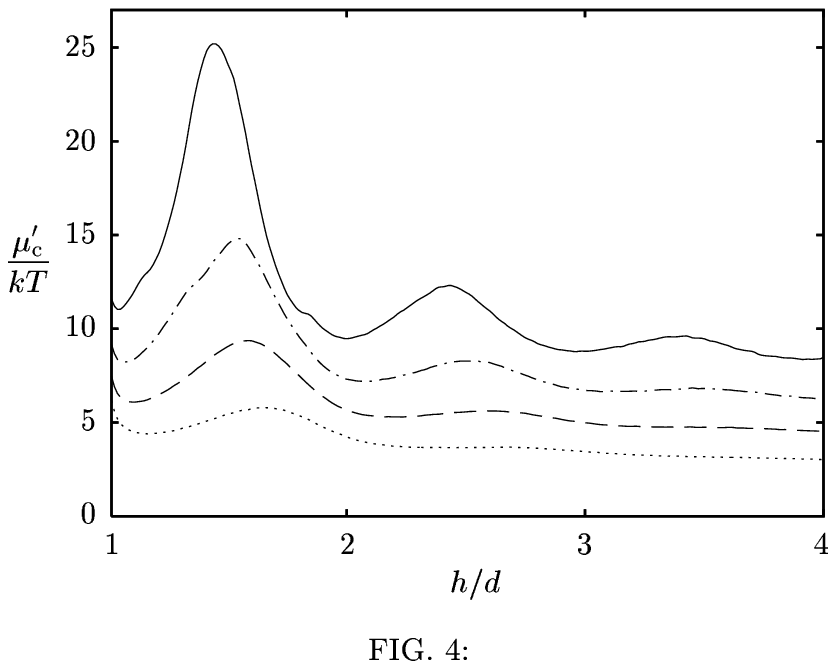}

\caption{
Excess chemical potential of the particles in the film, versus film
thickness $h$, for the same values of volume fraction
$\volumeFractionC$ as in Fig.\ \ref{pressures versus h}.  The
normalization corresponds to the asymptotic behavior
$\muCexcess=kT\ln(d^3\nCvolAverPrime)$ for $\volumeFractionC\ll1$.
}
\label{chemical potential versus h}
\end{figure}
\begin{figure}

\includegraphics*{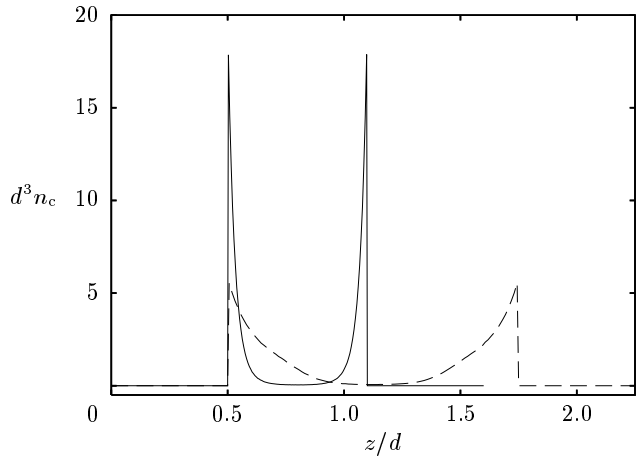}

\caption{
Local particle number density $\nCvol$ (per unit volume), versus
position $z$ across the film, for volume fraction
$\volumeFractionC=0.4$ and two values of film thickness $h/d=1.6$
(solid line) and $h/d=2.25$ (dashed line).  Relative to the film
thickness, the particle distribution is much narrower for $h/d=1.6$
(maximum of $\normalPC$ in Fig.\ \ref{pressures versus h}) than the
distribution for $h/d=2.25$ (minimum of $\normalPC$).
}
\label{local density}
\end{figure}

The results plotted in Fig.\ \ref{pressures versus h} indicate that
both $\normalPC$ and $\lateralPCaver$ are oscillatory functions of
$h$.  However, the phase of the oscillations of the normal and lateral
pressure components significantly differ, with the maxima of
$\lateralPCaver$ occurring at smaller values of $h$ than the maxima of
$\normalPC$.  Film tension \refeq{colloidal contribution to tension}
and chemical potential $\muCexcess$ [which is obtained from the
pressure tensor by integrating the Gibbs--Duhem relation
\refeq{modified excess Gibbs--Duhem relation}] are also oscillatory
functions of $h$.  (cf., Figs.\ \ref{film tension versus h} and
\ref{chemical potential versus h}).

The oscillatory behavior of the pressure tensor is typical of
effective structural interactions in confined domains.  The
nonmonotonic dependence on $h$ stems from the short-range layered
ordering of the suspension in the film
\cite{Vanderlick-Scriven-Davis:1989}.  A qualitative understanding of
the mechanism leading to the pressure oscillations and of the source
of the shift between the maxima of $\normalPC$ and $\lateralPCaver$
can be gained from an analysis of the suspension structure for
different values of the film thickness $h$.  In Fig.\ \ref{local
density} we show the local particle density $\nCvol$ in the film for
two values of interface separation, corresponding to the maximum
($h/d=1.6)$ and minimum ($h/d=2.25$) of the normal pressure
$\normalPC$ [cf., Fig.\ \ref{pressures versus h}\subfig{a}].

The results shown in Fig.\ \ref{local density} indicate that for both
values of $h/d$, there are two particle layers in the film.  For
$h/d=1.6$, the two layers barely fit into the available space $\hat
V$; particle concentration is thus strongly focused in thin regions
adjacent to each interface, and the suspension exerts high normal
pressure $\normalPC$ on the interfaces of the film.  For $h/d=2.25$,
the particle layers are more diffuse, and the normal pressure is much
lower.

In contrast to the behavior of $\normalPC$, the lateral pressure
component $\lateralPCaver$ remains moderate both for $h/d=1.6$ and
$h/d=2.25$ [cf.\ the results shown in Fig.\ \ref{pressures versus
h}\subfig{b}].  This is because for $h/d=1.6$ there are very few
particles in the central portion of the film, which implies that there
is no lateral momentum transport in this region.  The average value
thus remains relatively low, in spite of high momentum flux in the
layers of high particle concentration.  As a result of this mechanism,
the maxima of the lateral pressure $\lateralPC$ (averaged across the
film) occur for interface separations where both, the local lateral
momentum transport is relatively large, and it takes place over
sufficiently broad regions.

The oscillatory form of the intensive thermodynamic parameters
$\normalPC$, $\tensionC$, and $\muCexcess$ strongly affects properties
of particle-stabilized films.  First, owing to these oscillations, the
coexistence between uniform film phases of different thickness is
possible, as observed experimentally
\cite{%
Nikolov-Wasan-Kralchevsky-Ivanov:1992,%
Nikolov-Wasan:1992,%
Basheva-Danov-Kralchevsky:1997,%
Henderson-Trokhymchuk-Nikolov-Wasan:2005%
}
and discussed in our recent paper \cite{Blawzdziewicz-Wajnryb:2005}.

Next, a small perturbation $\delta h$ of the film thickness from a
uniform stable equilibrium state produces restoring force proportional
to the lateral gradient of $\delta h$.  In contrast, in particle-free
films the leading-order restoring force results from the capillary
pressure, and it is thus proportional to the gradient of the film
curvature $\curvature\sim\nablaS^2h$.  For long-wavelength
perturbations, the restoring force acting in particle-stabilized films
is thus significantly stronger than the restoring force in
particle-free films.  Therefore, particle-stabilized films often
retain nearly uniform thickness during the drainage process (except
for stepwise changes associated with phase transitions), whereas
particle-free films usually assume a dimpled shape with nearly
constant curvature.

\section{Macroscopic transport processes}
\label{Macroscopic transport properties}

After discussing equilibrium thermodynamics of particle-stabilized
thin liquid films, we now focus on linear transport processes that
occur when such films are perturbed from equilibrium.  In
Sec. \ref{Compressible-fluid analogy} we present quasi-two-dimensional
continuity equations for conserved parameters characterizing
local-equilibrium states of the film, and in Secs.\ \ref{Stress
balance equations} and \ref{Particle flux section} we formulate
constitutive relations between thermodynamic forces and macroscopic
fluxes in the conservation equations.  

\subsection{Conservation equations}
\label{Compressible-fluid analogy}

The thermodynamic analysis presented in Sec.\ \ref{Equilibrium
description} indicates that a particle-stabilized thin liquid film
behaves as a two-component 2D {\it compressible} fluid, even though
the suspension in the film is incompressible.  The film thickness
plays the role of the mass density, per unit area, and the number of
colloidal particles per unit area corresponds to the concentration of
the solute.

In the quasi-two-dimensional description, the macroscopic motion of
the suspension in the film is described by the lateral velocity
\begin{equation}
\label{lateral velocity}
\lateralVelocityVector=\lateralUnitTensor\bcdot\bar\bv,
\end{equation}
where $\bv$ denotes the ensemble-averaged velocity field, and the overbar
denotes the average across the film \refeq{average across film}.  The
macroscopic motion of the particles in the film is characterized by
the 2D lateral particle flux
\begin{equation}
\label{lateral particle flux}
\lateralFluxVector=\lateralUnitTensor\bcdot h\fluxVectorAver,
\end{equation}
where $\fluxVectorAver$ is the 3D particle flux $\fluxVector$ averaged
across the film.  We use the normalization where the particle flux
$\lateralFluxVector$ is measured with respect to the local suspension
velocity $\lateralVelocityVector$.

Definition \refeq{lateral velocity} and the incompressibility
condition \mbox{$\bnabla\bcdot\bv=0$} for the 3D flow field yield the
2D continuity equation for the film thickness $h$,
\begin{equation}
\label{2d continuity for flow}
\frac{\partial h}{\partial t}=-\bnablaS\bcdot(h\lateralVelocityVector),
\end{equation}
where $t$ denotes time.  Similarly, the evolution of the particle
density per unit area is governed by the 2D continuity equation
\begin{equation}
\label{2d continuity for particles}
\frac{\partial\nCarea}{\partial t}
  =-\bnablaS\bcdot(\nCarea\lateralVelocityVector+\lateralFluxVector).
\end{equation}

To obtain a full description of the film dynamics, the continuity
relations \refeq{2d continuity for flow} and \refeq{2d continuity for
particles} have to be supplemented with the stress-balance equations
and the constitutive equations for the stress tensor and the particle
flux in the film.

\subsection{Stress-balance equations}
\label{Stress balance equations}

\subsubsection{Lateral stress balance}

According to the results given in Appendix \ref{Effective lateral
stress}, the stress-balance equation in the film-tension
representation involves the effective 2D surface stress
$\surfaceStressTensor$ and area-force density \refeq{force per unit
surface},
\begin{equation}
\label{surface stress balance equation}
\bnablaS\bcdot\surfaceStressTensor=-\bFs.
\end{equation}
The effective surface stress, defined by relations \refeq{relative
stress} and \refeq{definition of effective surface stress}, equals the
excess stress beyond the constant isotropic background pressure
$\externalP\unitTensor$, integrated across the film.  As in all our
dynamic considerations, in the stress-balance equation \refeq{surface
stress balance equation}, creeping-flow conditions are assumed.

In equilibrium, the 2D surface-stress tensor reduces to the isotropic
film tension, $\surfaceStressTensor=\tension\lateralUnitTensor$.  In a
nonequilibrium state, the surface stress
\begin{equation}
\label{equilibrium and dissipative parts of surface stress}
\surfaceStressTensor=\tension\lateralUnitTensor
  +\viscousSurfaceStressTensor
\end{equation}
is a sum of the local-equilibrium contribution
$\tension\lateralUnitTensor$ and the viscous part
$\viscousSurfaceStressTensor$.  In the above equation, the local value
of the film tension $\tension$ is given by the equilibrium equation of
state, consistent with the standard local-equilibrium assumption
\cite{Mazur-de_Groot:1984}.  For a particle-stabilized film, the film
tension $\tension=\tension(h,\nCarea)$ is fully determined by the
local film thickness $h$ and local number of colloidal particles per
unit area of the film $\nCarea$.

In our further discussion, we assume the linear-response regime and
long-wavelength limit.  Under these conditions, the viscous stress
$\viscousSurfaceStressTensor$ is related to the lateral velocity
\refeq{lateral velocity} via two film-viscosity coefficients: the
shear and the expansion viscosities $\shearViscosityS$ and
$\expansionViscosityS$.  Accordingly, we have
\begin{equation}
\label{viscous stress}
\viscousSurfaceStressTensor
   =2\shearViscosityS\Deviatoric{\bnablaS\lateralVelocityVector}
   +\expansionViscosityS(\bnablaS\bcdot\lateralVelocityVector)
                                          \lateralUnitTensor,
\end{equation}
where
\begin{equation}
\label{deviatoric strain rate tensor}
\Deviatoric{\bnablaS\lateralVelocityVector}
  =\halff[\bnablaS\lateralVelocityVector
    +(\bnablaS\lateralVelocityVector)^\dagger
    -\bnablaS\bcdot\lateralVelocityVector\,\lateralUnitTensor]
\end{equation}
is the deviatoric part of the lateral strain-rate tensor
$\bnablaS\lateralVelocityVector$, with the dagger denoting the
transpose.

\subsubsection{Normal stress balance}

Owing to the suspension incompressibility, the normal balance of the
total stress in the film does not have an independent dynamical
meaning, unlike the lateral balance \refeq{surface stress balance
equation}.  However, the normal osmotic pressure component is an
important dynamic quantity, because it drives a lateral particle flux.
These results are derived below.

In a nonequilibrium state of the film, the equilibrium
condition \refeq{pressure equilibrium condition} is replaced by the
normal-stress balance
\begin{equation}
\label{nonequilibrium normal stress balance equation}
\normalStress=-\externalP,
\end{equation}
where $\normalStress$ is the normal component of the stress tensor
$\stressTensor$.  Unlike the effective surface stress
\refeq{equilibrium and dissipative parts of surface stress}, the
normal stress $\normalStress$ cannot be uniquely decomposed into the
equilibrium and nonequilibrium components.  This is because in an
incompressible suspension the isotropic component of the pressure
tensor (and thus also the normal pressure $\normalP$) is not a
well-defined function of the densities of the conserved quantities
$h=V/A$ and $\nCarea=\NC/A$.  There is thus no independent
constitutive relation for the nonequilibrium part of $\normalStress$,
and any unbalanced normal stress is immediately compensated for by a
proper redistribution of the isotropic component of the 3D pressure
tensor.  We note that such redistribution does not modify the lateral
stress balance in the film, because the surface stress
$\surfaceStressTensor$ involves only differences between the normal
and lateral components of the 3D stress, according to Eq.\
\refeq{definition of effective surface stress}. 

While the normal stress cannot be uniquely decomposed into the
equilibrium and nonequilibrium parts, there exists another useful
decomposition.  Namely, the stress $\normalStress$ can be decomposed
into the equilibrium osmotic normal pressure $\normalPC$ and the
remaining nonequilibrium part $\normalStressRest$,
\begin{equation}
\label{decomposition of normal pressure}
\normalStress=\normalPC+\normalStressRest.
\end{equation}
The nonequilibrium stress component $\normalStressRest$ involves the
3D isotropic fluid-pressure, and anisotropic contributions that
result from \subfig{a} the viscous fluid flow and \subfig{b} deviation
of the particle distribution from equilibrium.  Owing to the
equilibrium condition \refeq{nonequilibrium normal stress balance
equation}, for given $\externalP$ the stress $\normalStressRest$ is
entirely determined by the osmotic normal pressure $\normalPC$, which
is a well-defined function of $h$ and $\nCarea$.

\subsection{Particle flux}
\label{Particle flux section}

According to the principles of the thermodynamics of irreversible
processes \cite{Mazur-de_Groot:1984}, in the linear-response regime
the diffusive flux of colloidal particles is driven by gradients of
scalar intensive thermodynamic parameters that characterize the
thermodynamic equilibrium conditions.  In a particle-stabilized thin
liquid film, the relevant intensive parameters are the normal osmotic
pressure $\normalPC$ (as discussed above) and the chemical potential
of the suspended particles $\muCexcess$.  If a particle-stabilized
film is subject to external lateral forces, we also have thermodynamic
forces associated with the external potentials $\psiFluidVol$ and
$\psiCollExcess$, according to the equilibrium conditions
\refeq{equilibrium conditions for excess pressure} and
\refeq{equilibrium condition for excess chemical potential}.  However,
the spatial variations of the film tension $\tension$ (playing a
role analogous to the isotropic equilibrium pressure in simple fluid)
produce only the macroscopic motion of the film.  Hence, in the
linear-response regime, we assume the following constitutive relation
for the 2D diffusive particle flux
\begin{equation}
\label{constitutive relation for particle flux}
\lateralFluxVector=-\collectiveMobilityMu\bnablaS(\muCexcess+\psiCollExcess)
  -\collectiveMobilitySp\bnablaS(\normalPC-\psiFluidVol).
\end{equation}

Using the Gibbs--Duhem relation \refeq{excess Gibbs--Duhem relation}
it can be shown that the kinetic coefficients $\collectiveMobilityMu$
and $\collectiveMobilitySp$ cannot be determined separately.
Therefore, we combine them into a single mobility coefficient
$\collectiveMobilityExcess$.  We first consider a quiescent film with
no viscous stress \refeq{viscous stress}.  In this case the variations
of the osmotic contribution to film tension $\tensionC$ are related to
the external potentials $\psiCollExcess$ and $\psiFluidVol$ via the
hydrostatic condition
\begin{equation}
\label{hydrostatic condition for gamma in terms of potentials}
\bnablaS\tensionC=\nCarea\bnablaS\psiCollExcess +h\bnablaS\psiFluidVol,
\end{equation}
which follows from Eqs.\ \refeq{equilibrium condition for excess film
tension}, \refeq{volume potential acting on fluid}, and
\refeq{colloidal potential with buoyancy}.  From the Gibbs--Duhem
relation \refeq{excess Gibbs--Duhem relation} we also have
\begin{equation}
\label{Gibbs--Duhem for gradient of tension}
\bnablaS\tensionC=\nCarea\bnablaS\muCexcess-h\bnablaS\normalPC.
\end{equation}
Combining Eqs.\ \refeq{constitutive relation for particle flux} and
\refeq{hydrostatic condition for gamma in terms of potentials} yields
one-to-one correspondence between the thermodynamic forces on the
right-hand side of the constitutive relation \refeq{constitutive
relation for particle flux},
\begin{equation}
\label{relation between thermodynamic forces}
\nCvolAver\bnablaS(\muCexcess+\psiCollExcess)=\bnablaS(\normalPC-\psiFluidVol).
\end{equation}
Using \refeq{relation between thermodynamic forces} to eliminate the
osmotic normal pressure $\normalPC$ and potential $\psiFluidVol$ from
Eq.\ \refeq{constitutive relation for particle flux}, we find
\begin{equation}
\label{mobility relation}
\lateralFluxVector=-\nCarea\collectiveMobilityExcess
                    \bnablaS(\muCexcess+\psiCollExcess),
\end{equation}
where
\begin{equation}
\label{mobility coefficient}
\collectiveMobilityExcess=\nCarea^{-1}\collectiveMobilityMu
                   +h^{-1}\collectiveMobilitySp
\end{equation}
is the collective mobility of the colloidal particles in the film.  We
note that Eq.\ \refeq{mobility relation} relates the particle flux
$\lateralFluxVector$ to the {\it excess} quantities \refeq{excess
chemical potential} and \refeq{colloidal potential with buoyancy},
indicated by the prime in our notation for the mobility coefficient
\refeq{mobility coefficient} (cf.\ Sec.\ \ref{Particle mobility}).

In nonequilibrium states with nonvanishing gradients of the film
velocity $\lateralVelocityVector$, the balance equation for the
lateral stress involves the viscous contribution
$\bnablaS\bcdot\viscousSurfaceStressTensor$, according to equations
\refeq{surface stress balance equation} and \refeq{equilibrium and
dissipative parts of surface stress}.  As a result, a modified
relation \refeq{relation between thermodynamic forces} between the
thermodynamic forces in Eq.\ \refeq{constitutive relation for particle
flux} includes a $\bnablaS\bcdot\viscousSurfaceStressTensor$
correction.  This correction, however, is small in the long-wavelength
regime, because it depends on second derivatives of the film velocity
$\lateralVelocityVector$, according to the constitutive relation for
the stress \refeq{viscous stress}.  It follows that in the
linear-response and long-wavelength limit, particle transport in the
film can be described using the constitutive equation \refeq{mobility
relation} with a single mobility coefficient
$\collectiveMobilityExcess$.
 
\section{Evaluation of the short-time transport coefficients}
\label{Evaluation of short-time transport coefficients}

\subsection{Elementary hydrodynamic processes}
\label{Elementary hydrodynamic processes}

In this section we evaluate the effective film-viscosity coefficients
$\shearViscosityS$ and $\expansionViscosityS$ and the collective
particle mobility.  The film interfaces are assumed to be surfactant
free, so free-interface boundary conditions are assumed.

We focus on the short-time transport coefficients corresponding to the
initial system response to a startup of a flow or a sudden application
of an external force.  The short-time transport coefficients also
describe the linear response to high-frequency oscillatory forcing.

In the short-time (or high-frequency) regime the characteristic
timescale of the forcing is much shorter than the structural
relaxation time of the suspension in the film.  Thus the deviation of
the particle distribution from equilibrium is insignificant, and the
Brownian contribution to the transport coefficients can be neglected.
Accordingly, evaluation of the transport coefficients requires solving
appropriate many-particle hydrodynamic problems for the equilibrium
particle distribution of spheres in the film under creeping-flow
conditions.

\subsubsection{Shear viscosity of the film}
\label{Surface shear viscosity}

To evaluate the effective shear viscosity coefficient
$\shearViscosityS$, we consider a film undergoing the planar straining
flow with the strain-rate tensor
\begin{equation}
\label{strain-rate tensor}
\strainRateTensor=\halff[\bnabla\bv+(\bnabla\bv)^\dagger]
\end{equation}
given by 
\begin{equation}
\label{shear flow}
\strainRateTensor=
  \left[\begin{array}{rrr}
    1&0&0\\
    0&-1&0\\
    0&0&\phantom{-}0
  \end{array}\right],
\end{equation}
where the coordinates $x$ and $y$ are in the directions along the
film.  By the axial symmetry and translational invariance with respect
to lateral directions, the corresponding ensemble-averaged 3D
viscous-stress tensor has the form
\begin{equation}
\label{viscous stress tensor}
\viscousStressTensor=2\localShearViscosity(z)\strainRateTensor,
\end{equation}
where the mesoscopic viscosity coefficient $\localShearViscosity(z)$
depends on the transverse position in the film $z$.  Since equation
\refeq{viscous stress tensor} involves only the lateral components of
the stress, the effective shear viscosity $\shearViscosityS$ is
obtained by averaging Eq.\ \refeq{viscous stress tensor} across the
film.  Using notation \refeq{average across film} we thus get
\begin{equation}
\label{effective surface shear viscosity coefficient}
\shearViscosityS=h\averLocalShearViscosity,
\end{equation}
which follows from expressions \refeq{equilibrium and dissipative
parts of surface stress}, \refeq{viscous stress}, \refeq{relative
stress}, and \refeq{definition of effective surface stress}.

\begin{figure}

\includegraphics*{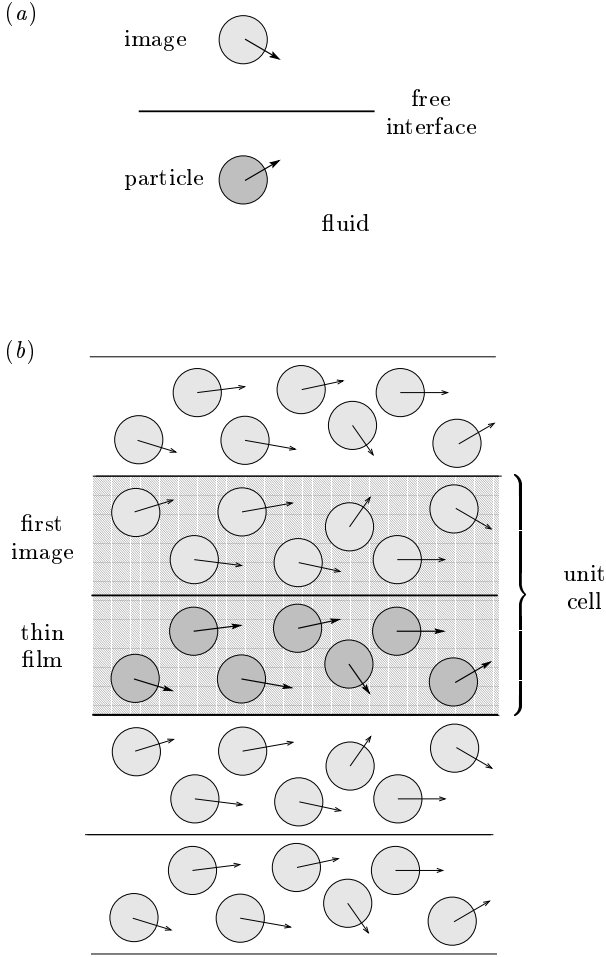}

\caption{%
Reflection method: \subfig{a} single free interface; \subfig{b}
representation of a particle-stabilized film in terms of an unbounded
periodic system.
}
\label{reflection method}
\end{figure}

\subsubsection{Expansion viscosity of the film}
\label{Surface expansion viscosity}

The effective expansion viscosity $\expansionViscosityS$ is obtained
by evaluating the stress response of the film to the axisymmetric
straining flow with the strain-rate tensor given by
\begin{equation}
\label{straining flow}
\strainRateTensor=
  \left[\begin{array}{rrr}
    1&0&0\\
    0&\phantom{-}1&0\\
    0&0&-2
  \end{array}\right],
\end{equation}
which corresponds to a uniform expansion of the film.  By axial
symmetry, the 3D viscous-stress tensor can be expressed
as a combination of the normal and lateral terms,
\begin{equation}
\label{stress for expansion problem}
\viscousStressTensor=\kappa_\perp\ez\ez+\kappa_\parallel(z)\lateralUnitTensor,
\end{equation}
where $\kappa_\perp$ is independent of $z$, owing to the continuity of
the transverse momentum flux.  Integrating the above relation with
respect to $z$, and recalling that the effective viscous stress is
associated with the excess part \refeq{definition of effective surface
stress} of the stress tensor, we find
\begin{equation}
\label{effective surface expansion viscosity coefficient}
\expansionViscosityS=h({\bar\kappa}_\parallel-\kappa_\perp).
\end{equation}

In our numerical calculations, the deviatoric part of the average
stress tensor \refeq{stress for expansion problem} is determined
directly from the stresslet component of the induced-force
distribution on the particle surfaces.  To obtain the relation that
links $\expansionViscosityS$ to our numerical results, we rewrite
equation \refeq{stress for expansion problem}, averaged across the
film, in the form
\begin{equation}
\label{average stress for expansion problem}
\viscousStressTensorAver=\bar\alpha\unitTensor
  +\bar\beta\strainRateTensor,
\end{equation}
where
\begin{equation}
\label{alpha and beta}
\bar\alpha=\third(\kappa_\perp+2{\bar\kappa}_\parallel),\qquad
\bar\beta=\third({\bar\kappa}_\parallel-\kappa_\perp),
\end{equation}
and $\strainRateTensor$ is the strain-rate tensor \refeq{straining
flow}.  Comparing relations \refeq{effective surface expansion
viscosity coefficient} and \refeq{alpha and beta}, we find that the
expansion film viscosity and the amplitude $\bar\beta$ of the
deviatoric part of the average stress \refeq{average stress for
expansion problem} are related by equation
\begin{equation}
\label{expansion viscosity in terms of deviatoric stress}
\expansionViscosityS=3h\bar\beta.
\end{equation}

\subsubsection{Particle mobility}
\label{Particle mobility}

The collective mobility coefficient \refeq{mobility coefficient} was
determined by evaluating the particle flux $\lateralFluxVector$
produced by a constant lateral external force \refeqa{two force fields
a,b}{b} in a uniform film with zero net force \refeq{force per unit
surface} and zero net volume flow.  The zero-net-force condition and
the definition \refeq{colloidal potential with buoyancy} of the excess
potential (i.e., potential corrected for the buoyancy contribution)
imply that the excess force and total force acting on the colloidal
particles are directly related,
\begin{equation}
\label{excess and not excess related}
\bnablaS\psiCollExcess=(1-\volumeFractionC)\bnablaS\psiColl.  
\end{equation}
Thus,
for a uniform system, Eq.\ \refeq{mobility relation} can be rewritten
as the mobility relation 
\begin{equation}
\label{definition of collective mobility non-excess}
\lateralFluxVector=-\nCarea\collectiveMobility\bnablaS\psiColl,
\end{equation}
where
\begin{equation}
\label{non-excess mobility coefficient}
\collectiveMobility=(1-\volumeFractionC)\collectiveMobilityExcess
\end{equation}
is the mobility coefficient describing the response of the system to
the total (rather than excess) force.  For compatibility with the
usual definition of the collective mobility in bulk suspensions, we
present our results for the coefficient $\collectiveMobility$ instead
of $\collectiveMobilityExcess$.

\subsection{Reflection method}
\label{Reflection method}

To obtain the shear and expansion film-viscosity coefficients
$\shearViscosityS$ and $\expansionViscosityS$, the stress response of
the film with a constant thickness $h$ to the flow fields \refeq{shear
flow} and \refeq{straining flow} has to be evaluated.  Similarly, the
collective mobility coefficient $\collectiveMobility$ is obtained by
calculating the particle flux produced by a constant lateral force
acting on the particles.  To solve these hydrodynamic problems we use
a standard Stokesian-dynamics algorithm for free space, combined with
a flow reflection method.

Our reflection technique is based on the well-known observation that
the flow reflected from a free interface at $z=0$ has the form
\begin{equation}
\label{flow reflected from free interface}
\vOut(x,y,z)=(\lateralUnitTensor-\ez\ez)\bcdot\vIn(x,y,-z),
\end{equation}
where $\vIn$ is the incoming flow.  Accordingly, the image of a
spherical particle centered at $(X,Y,Z)$ and moving with the velocity
$(U_x,U_y,U_z)$ is a sphere centered at $(X,Y,-Z)$, moving with the
velocity $(U_x,U_y,-U_z)$.  This reflection geometry is illustrated in
Fig.\ \ref{reflection method}\subfig{a}.

To obtain the solution for particle motion in a film with periodic
boundary conditions in the lateral directions, we use a
Stokesian-dynamics algorithm for a periodic system of particles in
free space.  We construct an elementary cell in 3D from an elementary
cell in the film and the reflection of this cell in the upper
interface, as illustrated in Fig.\ \ref{reflection method}\subfig{b}.
Multiple reflections of the composed elementary cell in the 3D system
ensure that the free-boundary boundary conditions are satisfied both
on the lower and the upper film interface.  Thus, the periodic 3D
system and the doubly periodic liquid film with free interfaces are
hydrodynamically equivalent.

We note, however, that our hydrodynamic calculations in the thin-film
geometry differ from the usual results for 3D periodic systems.
Namely, in the thin-film case, the spheres are not allowed to overlap
the lateral planes separating periodic cells, because these planes
represent the film interfaces.  This constraint is not required in
simulations of suspensions in an unbounded space.

In our numerical calculations we use the hydrodynamic-interactions
algorithm introduced in
\cite{Cichocki-Felderhof-Hinsen-Wajnryb-Blawzdziewicz:1994}, with
further improvements described in \cite{Ekiel_Jezewska-Wajnryb:2008}.
The algorithm combines a multipolar expansion of the flow field in the
system, with a superposition of two-particle lubrication corrections.
Our algorithm includes the multipolar expansion of Stokes flow to an
arbitrary order, thus accurate results, with controlled
multipolar-truncation error, can be obtained.

\begin{figure}

\includegraphics*{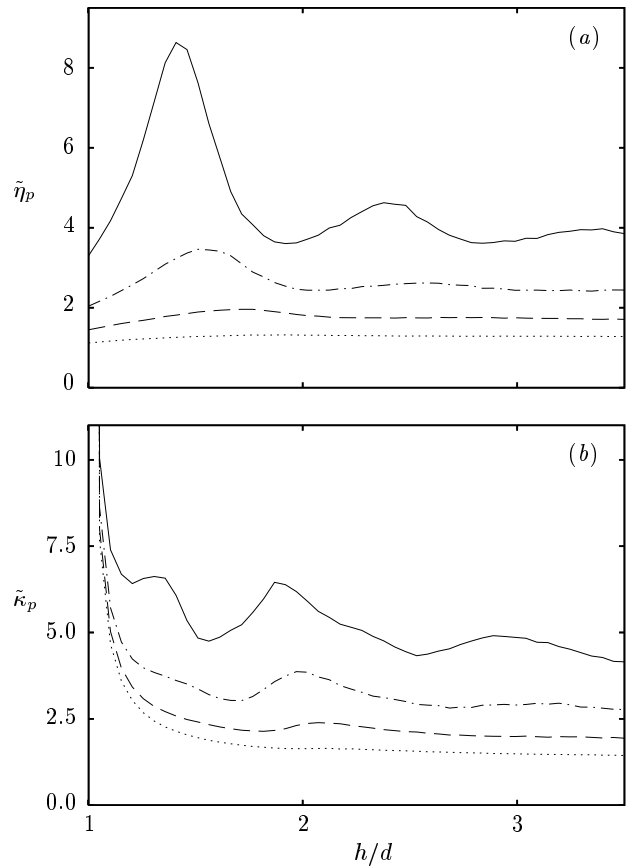}

\caption{
Normalized particle contributions to the \subfig{a} shear and
\subfig{b} expansion film-viscosity coefficients, as defined by Eqs.\
\refeq{particle contributions to viscosity}, versus film thickness
$h$, for particle volume fractions $\volumeFractionC=0.1$ (dotted
line), $0.2$ (dashed), $0.3$ (dash--doted), $0.4$ (solid).
}
\label{shear and compression viscosities}
\end{figure}


\subsection{Simulation details}
\label{Simulation details}

The short-time film viscosity coefficients $\shearViscosityS$ and
$\expansionViscosityS$ were determined using relation
\begin{equation}
\label{relation between stress and stresslet}
\viscousStressTensorAver=V^{-1}\sum_{i=1}^N \stresslet_i
\end{equation}
between the average stress tensor \refeq{effective surface shear
viscosity coefficient} or \refeq{average stress for expansion problem}
and the stresslets
\begin{equation}
\label{stresslet}
\stresslet_i=\frac{1}{2}\int 
  \left[
    \br_i\bF_i(\br_i)+\bF_i(\br_i)\br_i
      -\smallfrac{2}{3}\br_i\bcdot\bF_i(\br_i)\unitTensor
  \right]
  \,d\br_i
\end{equation}
of the induced-force distribution $\bF_i$ on the surfaces of particles
$i=1,\ldots,N$ (where $\br_i=\br-\bR_i$ is the relative position
vector with respect to the center of particle $i$).  The
film-viscosity coefficients $\shearViscosityS$ and
$\expansionViscosityS$ were evaluated using relations \refeq{effective
surface shear viscosity coefficient} and \refeq{expansion viscosity in
terms of deviatoric stress} for a suspension of spheres in the
external 2D straining flow \refeq{shear flow} and axisymmetric
straining \refeq{straining flow} flow, respectively.  The collective
lateral particle mobility was obtained from the relation
\begin{equation}
\label{collective mobility in terms of translational mobility coefficients}
\collectiveMobility
  =\halff N^{-1}\sum_{i,j=1}^N\mobilityMatrixTT_{ij}:\lateralUnitTensor,
\end{equation}
where $\mobilityMatrixTT_{ij}$ is the translational component of the
many-particle translational mobility matrix in a periodic system with
zero net flow.

The quantities \refeq{relation between stress and
stresslet}--\refeq{collective mobility in terms of translational
mobility coefficients} were determined for an equilibrium distribution
of $N=40,80$ and $160$ spheres in films of different thickness $h$.
The results were averaged over several hundred particle
configurations, obtained using a Monte--Carlo method.  In our
calculations, the force multipoles with the vector-spherical-harmonic
orders up to $l=5$ were included.

\begin{figure}

\includegraphics*{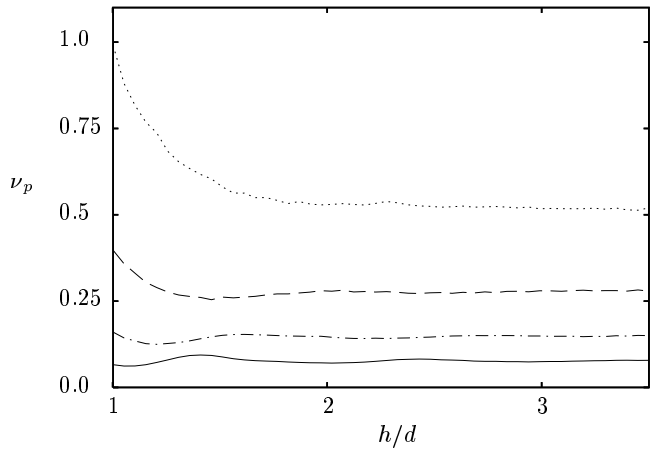}

\caption{
Normalized collective mobility coefficient, as defined by Eq.\
\refeq{normalized mobility}, versus film thickness $h$, for the same
values of volume fraction $\volumeFractionC$ as in Fig.\ \ref{shear
and compression viscosities}.
}
\label{collective mobility figure}
\end{figure}

\subsection{Numerical results}
\label{Numerical results}

Our numerical results for the shear and expansion film-viscosity
coefficients $\shearViscosityS$ and $\expansionViscosityS$ are
depicted in Fig.\ \ref{shear and compression viscosities}, and those
for the collective lateral mobility coefficient $\collectiveMobility$
are shown in Fig.\ \ref{collective mobility figure}.  The
film-viscosity coefficients $\shearViscosityS$ and
$\expansionViscosityS$ are presented in terms of the normalized
particle contributions $\shearViscositySC$ and
$\expansionViscositySC$, defined by the relations
\begin{subequations}
\label{particle contributions to viscosity}
\begin{equation}
\label{shear}
\shearViscosityS=\eta_0 h(1+\smallfrac{5}{2}\volumeFractionC\shearViscositySC),
\end{equation}
\begin{equation}
\label{extension}
\expansionViscosityS
   =\smallfrac{15}{2}\volumeFractionC\eta_0 h\expansionViscositySC,
\end{equation}
\end{subequations}
where $\eta_0$ is the viscosity of the suspending fluid.  Particle
mobility in the film is characterized by the dimensionless mobility
coefficient $\collectiveMobilityC$, normalized by the mobility of an
isolated particle in free space
\begin{equation}
\label{normalized mobility}
\collectiveMobility=3\pi\eta_0 d\collectiveMobilityC.
\end{equation}
With the above definitions, we have $\shearViscositySC,
\expansionViscositySC, \collectiveMobilityC\to1$ in the low-density
and thick-film limit $\volumeFractionC\to0$ and $h/d\to\infty$.  The
transport coefficients $\shearViscositySC$, $\expansionViscositySC$
and $\collectiveMobilityC$ are plotted in Figs.\ \ref{shear and
compression viscosities} and \ref{collective mobility figure} versus
the film thickness $h$ for several particle volume fractions
$\volumeFractionC$.

The results in Fig.\ \ref{shear and compression viscosities} indicate
that the normalized film-viscosity coefficients depend on the film
thickness in an oscillatory way, especially in the high
volume-fraction regime.  The oscillations are more pronounced for the
shear-viscosity coefficient, which has a large peak at $h/d\approx1.4$
[close to the position of the peak of the lateral pressure
$\lateralPCaver$, shown in Fig.\ \ref{pressures versus h}\subfig{b}].
At high volume fractions, the normalized mobility coefficient
$\collectiveMobilityC$, depicted in Fig.\ \ref{collective mobility
figure}, also has significant oscillations.  The oscillations are more
clearly visible in the plot shown in Fig.\ \ref{L
convergence}\subfig{a} on an expanded scale.

According to the results depicted in Fig.\ \ref{shear and compression
viscosities}\subfig{a}, the normalized shear viscosity of the film
$\shearViscositySC$ assumes a minimal value at $h/d=1$.  This behavior
stems from the fact that for lateral particle motion, the image
particles move in the same direction as the original ones (cf.\ the
schematic representation in Fig.\ \ref{reflection method}).  Such a
collective motion results in a reduced dissipation.  For the same
reason, the collective mobility $\collectiveMobilityC$ has a maximum
at $h/d=1$, as shown in Fig.\ \ref{collective mobility figure}.  In
contrast, the expansion-viscosity coefficient $\expansionViscositySC$,
plotted in Fig.\ \ref{shear and compression viscosities}\subfig{b},
has an $O(\epsilon^{-1})$ singularity at $h/d=1$ (where
$\epsilon=h/d-1$ denotes a typical gap between the particle surfaces
and the film interfaces), owing to the lubrication resistance between
the original particles and their images (which have the opposite
normal velocity component).  At low particle volume fractions the
collective mobility coefficient $\collectiveMobilityC$ diverges as
$\log(\volumeFractionC)$, because of the logarithmic far-field
behavior of 2D Stokes flow.

\begin{figure}

\includegraphics*{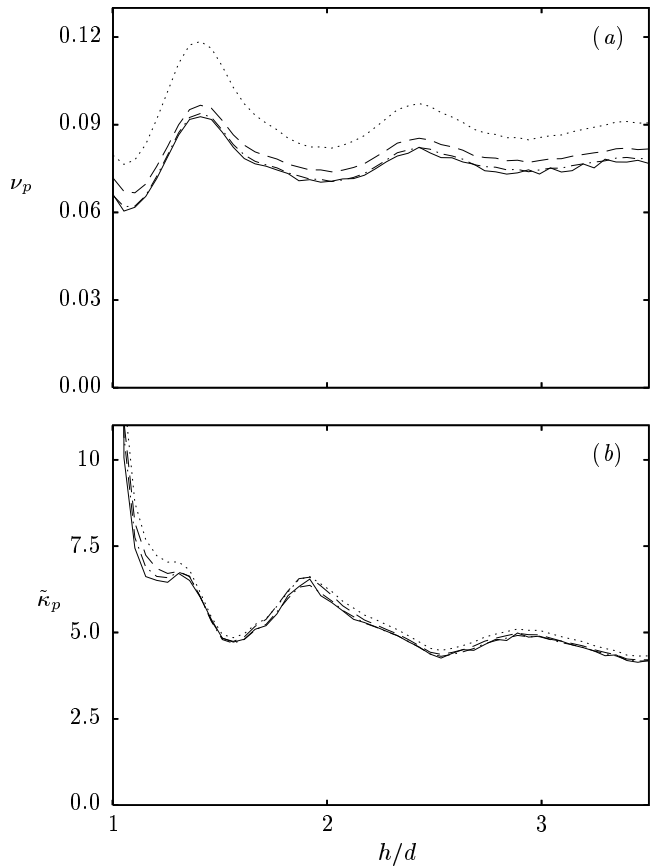}

\caption{
Normalized \subfig{a} collective mobility coefficient and \subfig{b}
particle contribution to the expansion viscosity of a
particle-stabilized film, versus film thickness $h$, for
$\volumeFractionC=0.4$ and spherical-harmonics truncation order
$\lmax=2$ (dotted line), 3 (dashed) 4 (dash--dotted), and 5 (solid).
}
\label{L convergence}
\end{figure}
\begin{figure}

\includegraphics*{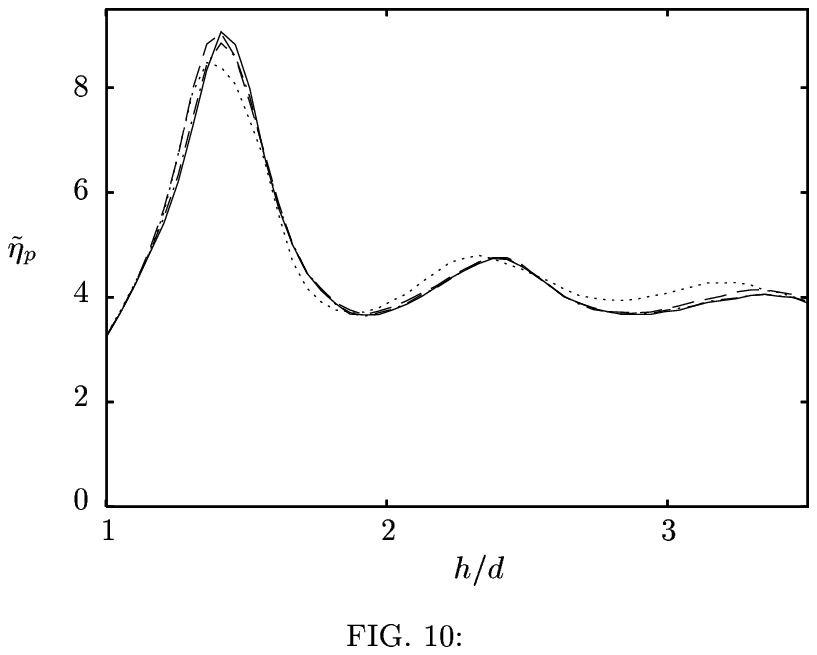}

\caption{
Normalized particle contributions to the shear viscosity of a
particle-stabilized film, versus the film thickness $h$, for
$\volumeFractionC=0.4$ and system size $N=20$ particles (dotted line),
40 (dashed), 80 (dashed--dotted), and 160 (solid).
}
\label{N convergence}
\end{figure}

\subsection{Convergence tests}
\label{Convergence tests}

The results for the film-viscosity and mobility coefficients shown in
Figs.\ \ref{shear and compression viscosities} and \ref{collective
mobility figure} were obtained for a system of $N=80$ particles using
induced-force multipoles up to the spherical-harmonics order
$\lmax=5$.  The convergence of the results with $\lmax$ is illustrated
in Fig.\ \ref{L convergence} and with the system size $N$ in Fig.\
\ref{N convergence} for the highest particle volume fraction
considered in our simulations, $\volumeFractionC=0.4$.

The results shown in Fig.\ \ref{L convergence}\subfig{a} for the
mobility coefficient $\collectiveMobility$ indicate that the
truncation of the multipolar-expansion at the order $\lmax=2$ yields
an error of 20\,\% to 30\,\% for this quantity.  The results obtained
with $\lmax=3$ have the accuracy better than 8\,\% and $\lmax=4$
yields results accurate within $1\,\%$.  We note that $\lmax=2$
includes all multipoles of the STD scheme used by Brady and his
collaborators \cite{Brady-Bossis:1988}, and thirteen other flow
multipoles (including potential flows). Thus, for the mobility
coefficient $\collectiveMobility$ the standard STD scheme is
insufficient for obtaining accurate results, even if the lubrication
corrections are included in the numerical scheme.

Our numerical results for the film-viscosity coefficients
$\shearViscosityS$ and $\expansionViscosityS$ indicate that the
convergence with $\lmax$ is faster for these quantities than the
convergence for the collective mobility $\collectiveMobility$.  As
illustrated in Fig.\ \ref{L convergence}\subfig{b}, the truncation at
$\lmax=2$ already yields quite accurate results for the coefficient
$\expansionViscosityS$, except for films with $h/d\alt1.5$.  For
$\shearViscosityS$ the convergence with $\lmax$ is similar.

The convergence of our results with the system size is illustrated in
Fig.\ \ref{N convergence}, where the results for the effective
shear-viscosity coefficient $\shearViscosityS$, evaluated with
$\lmax=2$, are shown for systems with $N$ between 20 and 160 particles
in a unit cell.  The results indicate that the convergence of the
shear viscosity $\shearViscosityS$ is fast.  A similar convergence has
been obtained obtained for the expansion viscosity
$\expansionViscosityS$ and the collective mobility
$\collectiveMobility$.

\section{Conclusions}
\label{Conclusions}

Our unified quasi-two-dimensional description of equilibrium states
and transport properties in particle-stabilized thin liquid films
relies on the film-tension representation of the film thermodynamics.
In this representation the system is characterized by extensive
parameters corresponding to conserved quantities: the film volume,
film area, and the number of fluid molecules and colloidal particles.
The conjugate intensive parameters are the normal pressure, film
tension, and chemical potentials of the fluid phase and the colloidal
component.

Equilibrium conditions and their physical interpretation have been
analyzed for nonuniform films in external potentials.  In particular,
we have shown that the particle contribution to film tension is
associated with the excess lateral pressure beyond the uniform
isotropic background pressure $\externalP$.  Since the normal pressure
in the film is equal to the external pressure, the excess lateral
pressure equals the difference between the lateral and normal pressure
components.

Based on our thermodynamic formalism and the local-equilibrium
assumption, we have also introduced a quasi-two-dimensional
description of transport processes in particle-stabilized liquid
films.  In some aspects, a film composed of an {\it incompressible}
suspension is analogous to a 2D {\it compressible} fluid, with the
film thickness playing the role of the mass density per unit area, and
film tension being similar to the pressure (with a minus sign).

When deviations from equilibrium are small, the momentum transport in
the film is characterized by two linear viscosity coefficients: the
shear and expansion viscosities.  The particle diffusive transport is
described by two kinetic coefficients relating the particle flux to
the gradients of the normal component of the osmotic pressure
$\normalPC$ and the excess particle chemical potential $\muCexcess$.
Using the Gibbs-Duhem relation in the film-tension representation, the
two kinetic coefficients have been combined into a single collective
mobility coefficient.  Our explicit calculations of the short-time
linear transport coefficients for a film stabilized by a hard-sphere
suspension indicate that these coefficients are oscillatory functions
of film thickness.  The oscillations are due to the short-range
layering order in the film.

While the results of our investigations provide a convenient framework
for describing equilibrium and nonequilibrium properties of
particle-stabilized thin liquid films, there still remain many open
questions.  First, we considered only films with surfactant-free
interfaces.  Surfactant adsorbed on the interfaces would contribute to
film tension, but otherwise it should not significantly alter the
equilibrium conditions in the film.  It would, however, substantially
influence the relaxation times for decay of nonuniformities of
different intensive thermodynamic parameters.  Thus partial
equilibrium states in stratified films (discussed in
\cite{Blawzdziewicz-Wajnryb:2005}) can be significantly affected.

Adsorbed surfactant modifies not only the equilibrium equations of
state, but also changes the film viscosity and particle mobility,
because of its effect on the boundary conditions at the film
interfaces (c.f., our analysis of this problem presented in
\cite{Blawzdziewicz-Cristini-Loewenberg:1999}).  The
hydrodynamic-interaction algorithm used in the present paper is
applicable only to surfactant-free interfaces.  However, our recently
developed Cartesian-representation method
\cite{%
Bhattacharya-Blawzdziewicz-Wajnryb:2005a,%
Bhattacharya-Blawzdziewicz-Wajnryb:2005,%
Bhattacharya-Blawzdziewicz-Wajnryb:2006%
})
can be applied to evaluate the effective transport coefficients in a
particle-stabilized film with more complex boundary conditions.

Another important problem that has not been addressed here is
transport through contact lines between phases of different thickness
in stratified films.  An analysis in
\cite{Kralchevsky-Nikolov-Wasan-Ivanov:1990} suggests that such
transport may control the rate of expansion of thin spots in the film
during a stepwise-thinning process.  On the other hand, hydrodynamic
scaling arguments \cite{Blawzdziewicz-Wajnryb:2005} seem to imply that
particle transport through a contact line should be faster than
particle diffusion (at least in surfactant-free films).  This
transport, however, can be significantly slowed down, owing to
surfactant effects, and because of the entropic barrier associated
with a change of particle arrangement at the line separating phases of
different thickness.  This problem will be analyzed in our future
publications.

\acknowledgments{This work was supported by NSF CAREER grant CTS-0348175;
EW was also supported by Polish Ministry of Science grant N501 020 32/1994.}

\appendix

\section{Effective surface stress}
\label{Effective lateral stress}

In this appendix we derive the effective lateral stress-balance
equation under creeping-flow conditions in a thin film with varying
width $h$.  The film occupies the region between the lower and upper
interfaces
\begin{equation}
\label{position of interfaces}
z=\zlow(x,y),\qquad z=\zup(x,y).
\end{equation}
The region occupied by the film can be described by the characteristic
function,
\begin{equation}
\label{characteristic function of the film}
\tauFilm(\br)=\theta(z-\zlow)\theta(\zup-z)
\end{equation}
where $\theta$ denotes the Heaviside step function.  

The film is surrounded by a gas with a constant pressure $\externalP$,
and a lateral force field
\begin{equation}
\label{force on the film}
\bff(\br)=\tauFilm(\br)\bff(\br)
\end{equation}
acts on the film.  Accordingly, the stress tensor in the whole 3D
space
\begin{equation}
\label{stress tensor in the system}
\stressTensorWhole=-(1-\tauFilm)\externalP\unitTensor+\tauFilm\stressTensor
\end{equation}
(where $\stressTensor$ denotes the stress tensor in the film, and
$\externalP$ is the external pressure) satisfies the full 3D balance
equation
\begin{equation}
\label{stress balance in whole space}
\bnabla\bcdot\stressTensorWhole=-\bff.
\end{equation}

There are no capillary-pressure and Marangoni-stress contributions in
the above equation, because of our assumptions that the curvature of
the interfaces is small, and the interfacial tension is constant.  If
needed, such contributions can be added as appropriate delta
distributions of the force on the film interfaces.

To obtain the film-tension representation for the lateral stress
balance in the film, it is convenient to reorder terms in Eq.\
\refeq{stress tensor in the system},
\begin{equation}
\label{stress tensor in the system reordered}
\stressTensorWhole=-\externalP\unitTensor+\tauFilm\delta\stressTensor.
\end{equation}
The first term on the right-hand side of this equation is the constant
background pressure that does not contribute to the stress-balance
\refeq{stress balance in whole space}.  The second term represents the
excess stress in the film.  Using the continuity of the transverse
momentum flux and the associated boundary condition
\refeq{nonequilibrium normal stress balance equation} for the normal
stress we find
\begin{equation}
\label{relative stress}
\delta\stressTensor=(\lateralStress-\normalStress)\lateralUnitTensor.
\end{equation}
By inserting \refeq{stress tensor in the system reordered} into the
stress-balance equation \refeq{stress balance in whole space} and
integrating with respect to the variable $z$, we obtain
\begin{equation}
\label{intermediate step for lateral stres balance}
\int_{\infty}^{\infty}\,dz\,
   \left(\ez\frac{\partial}{\partial z}+\bnablaS\right)
   \bcdot\tauFilm\delta\stressTensor=-\bFs,
\end{equation}
where $\bFs$ is the force density per unit area of the film.  The
first term on the left-hand side of the above relation vanishes
because $\tauFilm(x,y,z)=0$ for $z=\pm\infty$.  In the second term we
change the order of integration and differentiation to obtain the
quasi-two-dimensional stress-balance equation for the excess stress,
\begin{equation}
\label{integrated lateral stress balance equation}
\bnablaS\bcdot h\delta\bar\stressTensor=-\bFs,
\end{equation}
where the bar denotes the average \refeq{average across film}.

In an equilibrium state, only the equilibrium pressure
tensor \refeq{pressure tensor} contributes to $\stressTensor$,
\begin{equation}
\label{equilibrium stress tensor}
\stressTensor=-(\normalP\ez\ez+\lateralP\lateralUnitTensor).
\end{equation}
Taking into account the boundary condition \refeq{pressure equilibrium
condition} and the definition of film tension \refeq{film tension in
terms of stress components}, we find that relations \refeq{integrated
lateral stress balance equation} and \refeq{equilibrium stress tensor}
yield the lateral stress balance \refeq{tension balance}.

In a more general, nonequilibrium case, Eq.\ \refeq{integrated lateral
stress balance equation} is split into the lateral and vertical
components,
\begin{equation}
\label{lateral component of lateral stress balance}
\bnablaS\bcdot h\delta\bar\stressTensor\bcdot\lateralUnitTensor=-\bFs.
\end{equation}
\begin{equation}
\label{normal component of lateral stress balance}
\bnablaS\bcdot h\delta\bar\stressTensor\bcdot\ez=0,
\end{equation}
Since the normal component \refeq{normal component of lateral stress
balance} vanishes on the assumption that there is no transverse
external force, we find that \refeq{lateral component of lateral
stress balance} yields the lateral stress balance \refeq{surface
stress balance equation}, with the effective 2D stress
\begin{equation}
\label{definition of effective surface stress}
\surfaceStressTensor=h\delta\bar\stressTensor\bcdot\lateralUnitTensor.
\end{equation}

\bibliographystyle{apsrev} 

\begin{thebibliography}{36}
\bibitem[{col()}]{collision_note}
\bibinfo{note}{Momentum transfer in suspensions is mediated through the
  suspending fluid. However, equilibrium properties of a suspension of hard
  spheres in an inert incompressible medium are equivalent to those of a fluid
  of hard spheres undergoing Newtonian dynamics in empty space. Therefore, in
  equilibrium considerations we can entirely neglect the effect of the
  suspending fluid.}
\end{thebibliography}

\end{document}